\documentstyle[aps,amssymb]{revtex}
\def\Bbb{\mathbb}
\def\BZ{{\Bbb Z}} \def\BC{{\Bbb C}} \def\BP{{\Bbb P}} \def\BR{{\Bbb R}}
\begin{document}
\title{Orientifolds of type IIA strings on
Calabi-Yau manifolds\footnote{Based on a talk presented by S.G. at PASCOS
2003 held at the Tata Institute of Fundamental Research, Mumbai during Jan.
3-8, 2003.} } 
\author{Suresh Govindarajan$^{\circ}$ and Jaydeep Majumder$^{\dagger}$} 
\address{$^{\circ}$Department of Physics, 
Indian Institute of Technology Madras, Chennai 600 036 India\\
$^{\dagger}$
Department of Physics and Astronomy, Rutgers University,
 Piscataway, NJ 08855-0849 USA} 
\maketitle
\begin{abstract}
We identify type IIA orientifolds that are dual to M-theory
compactifications on manifolds with $G_2$-holonomy. 
We then discuss the construction of crosscap states in Gepner models.
\end{abstract}
%\section{Introduction}
\medskip

The advent of D-branes has lead to a better understanding of dualities
involving strong coupling limits. In particular, ${\cal N}=1$
compactifications of the
heterotic string (on Calabi-Yau manifolds) are no longer the only string
theories of phenomenological interest. One such class is furnished by
M-theory compactifications on seven-dimensional 
manifolds of $G_2$-holonomy give rise to
a four-dimensional theories with ${\cal N}=1$  supersymmetry. When the
$G_2$ manifolds have certain kinds of singularities, both non-abelian
gauge groups as well as chiral fermions can appear.

Joyce has constructed manifolds of $G_2$ holonomy as $\BZ_{2}$ 
orbifolds of a Calabi-Yau threefold $M$:
$X=(M\times S^1)/\sigma\cdot{\cal I}_1$ where $\sigma$
is an anti-holomorphic involution of the $CY^3$ and ${\cal I}_1$,
inversion of the $S^1$\cite{Joyce}. 
One obtains a smooth manifold when the orbifold
action has no fixed points.
However, when there is a fixed point set $\Sigma$, one obtains a singular 
manifold\cite{HM,PP}.  
The singularity can been smoothed out when $b_1(\Sigma)>0$.
The focus of this talk will be on the cases when there are fixed points.

Our working example of a Joyce manifold is the one
obtained from the Fermat quintic given by the hypersurface 
$ z_1^5 + z_2^5 + z_3^5 + z_4^5 + z_5^5 =0$,
in $\BC\BP^{4}$($z_i$ are homogeneous coordinates of $\BC\BP^{4}$).
The anti-holomorphic involution $\sigma$ is
$z_i\rightarrow \bar{z}_i$ for $i=1,\ldots,5$. The
fixed-point set $\Sigma$
is an $\BR\BP^3$, which is a special Lagrangian(sL) submanifold of the 
Fermat quintic\cite{BBS}.
Since $b_1(\BR\BP^3)=0$, the singularity of $X$, which is
locally of the form $\Sigma\times \BC^2/\BZ_2$, {\em cannot} be resolved.
$\Sigma$ is actually one in a family of $5^4=625$ sL submanifolds of the
Fermat quintic, all of whom are $\BR\BP^3$'s. They are all fixed-points
of the anti-holomorphic involutions:
$z_i \rightarrow \alpha^{n_i}\ \bar{z}_i$ with $\alpha^5=1$.

We will focus on obtaining the precise type IIA orientifold dual for M-theory
compactification on this Joyce manifold. We then will proceed to study
the orientifolding in the Gepner model corresponding to the Fermat quintic.
This involves the construction of crosscap states in Gepner models which
we schematically discuss postponing details to a subsequent paper\cite{SGJM}.\\

\leftline{\bf Obtaining the orientifold dual}
\smallskip

M-theory compactified on $M\times S^1$ is dual to the type IIA
compactification on $M$. Since the Joyce manifold $X$ is an orbifold
of $M\times S^1$, the type IIA dual can be obtained if
we can identify the action of ${\cal I}_1$ on the type IIA
side. 
But, ${\cal I}_1$ is not a symmetry of M-theory and thus
cannot quite be identified with a symmetry on the type IIA side.
However, the inversion of
an even number of coordinates is a symmetry of M-theory. 
In the example of the quintic that we considered,
$\Sigma$ is the base of
the SYZ $T^3$ fibration of the quintic and $\sigma$ inverts the fibre\cite{SYZ}. 
Thus, $\sigma\cdot{\cal I}_1$ corresponds
to the simultaneous inversion of four circles -- three from the SYZ fibre
and one from the $S^1$.
This uniquely fixes the type IIA orientifold to be the second choice
from the following two possibilities\cite{Ashoke,Jaydeep}:
($\Omega$: worldsheet parity, $F_L$: spacetime fermion number)
$$
[\sigma\cdot \Omega]\qquad {\rm or} \qquad[(-)^{F_L}\cdot\sigma\cdot \Omega ]
\ .
$$
It also turns out only the second choice preserves ${\cal N}=1$
supersymmetry\cite{GM,KM}. This is easily
understood by studying the action on the vertex operators involving
the Ramond sector. 

The spectrum of M-theory on
$M\times S^1$ has ${\cal N}=2$ supersymmetry in $d=4$ and consists 
of: (a) the ${\cal N}=2$ supergravity multiplet; 
(b) $h_{1,1}(M)$ abelian vector multiplets; and 
(c) $h_{2,1}(M)+1$ hypermultiplets. 
The orbifolding breaks half the supersymmetry and 
the spectrum for a smooth Joyce manifold $X$ (with Betti numbers $b_3$ and
$b_2$) consists of\cite{PT,VW}:
(a) the ${\cal N}=1$ supergravity multiplet; 
(b) $b_2(X)=h_{1,1}^+(M)$
 abelian vector multiplets; and 
(c) $b_3(X)=h_{2,1}(M)+h_{1,1}^-(M)+1$ chiral multiplets, 
where $h_{1,1}^{\pm}(M)$ 
are the number of Kahler moduli that are even[odd] under $\sigma$.
For the case when the orbifolding has fixed points, additional moduli
appear corresponding to modes that smoothen the singularity.

For the Fermat quintic, 
$h_{2,1}=101\ ;\ h_{1,1}^+=0\ ;\ h_{1,1}^-=1$ and the singularity cannot
be resolved. The
two fixed points are of the form $\Sigma \times \BR^{3,1}$
and the singularity is locally like $\BR^4/\BZ_2$,
i.e., it is an $A_1$ singularity -- expect $U(1)\times U(1)$ enhanced
gauge symmetry in M-theory.
In the type IIA dual, we expect an $O6$-plane with
the $SO$-projection.
The RR-charge will be equal the $\BR^3/\BZ_2$
orientifold plane in flat space. Based on this, we add 4 $D6$-branes
wrapping $\Sigma \times \BR^{3,1}$ implying a $SO(4)$ gauge symmetry.
The rest of the talk will be towards checking if this geometric intuition
can be realised in the orientifold of the Gepner model associated with
the Fermat quintic.\\

\leftline{\bf Aspects of Orientifolding}\smallskip

It is useful to understand how the M-theory spectrum on $X$ must appear
from the orientifold projection in the type IIA theory on $M$. 
Let $\widetilde{\Omega}$ denote the orientifolding $\BZ_2$($=(-)^{F_L}
\cdot\sigma\cdot\Omega$ in our example).
Under its action, the states
of the original type IIA theory fall into three representations (which
we label by $\epsilon~=~0,\pm1$):
{\em Real representations}:[$\epsilon=+1$] 
These have eigenvalue $+1$ and survive orientifold projection.
{\em Pseudo-real  representations:}[$\epsilon=-1$] 
These have eigenvalue $-1$. and are projected out.
{\em Complex representations:}[$\epsilon=0$]  
Under the action of $\widetilde{\Omega}$, one state gets mapped to another. 
In such cases, one linear combination is projected out.
In our example, it is easy to see that states that arise from the $(c,c)$
and $(a,a)$ rings (complex moduli of $M$) 
are in the complex representation while those
that arise from the $(a,c)$ and $(c,a)$ rings (K\"ahler moduli of $M$)
have $\epsilon=\pm1$.

The presence of orientifold planes leads to unoriented strings and hence
unoriented surfaces. At `one-loop', this adds a Klein bottle to the
torus.
The Klein bottle amplitude has two ``channels'' related by the modular
transformation. The {\em direct channel} 
$K(q)= {\rm Tr}\left(\widetilde{\Omega}\ q^{H_{cl}}\right)
=\sum_i \epsilon_i\ \chi_i(q)$ 
and the {\em transverse channel}
$\tilde{K}(\tilde{q})=\langle C 
| \tilde{q}^{H_{cl}}|C\rangle 
=\sum_j {\Gamma_i}^2\ \chi_i(\tilde{q})$ 
%$\epsilon_i=0,\pm1$ 
%&&$|C\rangle=$ crosscap state\\ \hline
We have assumed for simplicity that all states have multiplicity of one. 
Thus, the direct channel amplitude encodes the orientifold projection.

In the CFT of unoriented strings, one first constructs a crosscap state
whose direct channel amplitude encodes the required projection. One
general class of solutions has been provided by 
Pradisi-Sagnotti-Stanev\cite{PSS}. The crosscap state is 
%\\[3pt]
%\centerline{
$$ |C\rangle = \sum_i \Gamma_i\ |C:i\rangle\rangle =
\sum_i \frac{P_{0i}}{\sqrt{S_{0i}}}\ |C:i\rangle\rangle\ , $$
%}
%\smallskip
where $|C:i\rangle\rangle$ are the Ishibashi basis for crosscap states
and $P\equiv \sqrt{T} S T^2 S \sqrt{T}$. This plays the analogue of
the S-matrix in Cardy's ansatz for the boundary states.
The matrices
$ Y_{ij}^k \equiv \sum_m \frac{S_{mi}P_{mj}P_m^k}{S_{m0}}$
plays a role analogous to the fusion matrix for boundary states.
They satisfy the fusion algebra: $Y_i\ Y_j = {N_{ij}}^k\ Y_k$ with
$Y_{00}^k=\epsilon_k$ determining the KB projection. \\
\leftline{\bf An application: ${\cal N}=2$ minimal models}\smallskip

The states in the minimal model of level $k$
 are labeled by $(L,M,S)$ with $L=0,\ldots,k$, $M=0,\ldots (2k+3)$ 
mod $(2k+4)$, $S=0,1,2,3$ mod $4$ and $L+M+S=even$. There is an additional
identification: $(L,M,S)\sim (k-L,M+k+2,S+2)$. Even $S$ is the NS-sector
and odd $S$ is the R-sector.
The S-matrix and P-matrix are schematically given by
\begin{eqnarray*}
S_{LMS}^{\tilde{L}\tilde{M}\tilde{S}} &\propto&
\sin (L,\tilde{L})_k\ e^{\frac{i\pi M\tilde{M}}{k+2}}\ 
e^{\frac{-i\pi S\tilde{S}}{2}}\\
P_{LMS}^{\tilde{L}\tilde{M}\tilde{S}} &\propto&\left(
\sin \frac12(L,\tilde{L})_k\ e^{\frac{i\pi M\tilde{M}}{(2k+4)}}\ 
e^{\frac{-i\pi S\tilde{S}}{4}}\ \delta^{(2)}_{M+\tilde{M}+k} \ 
\delta^{(2)}_{S+\tilde{S}} \right.\\
&&\left. +e^{i\alpha_{LMS}}\ \sin \frac12(k-L,\tilde{L})_k\ e^{\frac{i\pi (M+k+2)\tilde{M}}{(2k+4)}}\ 
e^{\frac{-i\pi (S+2)\tilde{S}}{4}}\ \delta^{(2)}_{M+\tilde{M}} \ 
\delta^{(2)}_{S+\tilde{S}}\right)
\end{eqnarray*}
where $(L,\tilde{L})_k= \pi(L+1)(\tilde{L}+1)/(k+2)$ and $\alpha_{LMS}$ is a
phase that one needs to introduce to take care of the identification
mentioned earlier\cite{Hikida,BH,SGJM}. 
The appearance of a Kronecker delta function in P-matrix implies that
only NSNS (or RR)states alone appear in the PSS crosscap state. \\
\leftline{\bf Crosscap states in the Gepner model}\smallskip

The Gepner model is obtained by tensoring
copies of ${\cal N}=2$ minimal models(MM) such that total central charge
is 9. For the quintic -- tensor five copies of $k=3$ MM.
Further, restrict to states that come from
tensoring NS states with NS states and R with R from each 
minimal model and project onto states with total (including spacetime sector)
$U(1)$ charge an odd integer.

This suggests the following strategy for crosscap states in the Gepner
model:
Take the tensor product of crosscap states in the individual minimal
model.
Implement the Gepner projection on this
crosscap state.
This is a natural guess for the crosscap state in the
Gepner model. 
But this cannot be the crosscap that realises the type IIA orientifold!
This is because PSS crosscap state has only contributions from the NSNS sector.
This implies that its Ramond charge is zero.
The direct channel KB amplitude is not  supersymmetric.

Consider the two crosscap states in a single MM
$$
|C:NSNS\rangle \equiv P_{000}^{LMS}\ |C:LMS\rangle\rangle  \quad {\rm and}\quad
|C:RR\rangle \equiv P_{011}^{LMS}\ |C:LMS\rangle\rangle 
$$
The first one is the PSS crosscap state 
while the second one is the PSS crosscap state associated with the
simple current that is related to spacetime supersymmetry. It contains
only RR Ishibashi states.
Then, we propose that the correct crosscap state schematically
takes the form
$$
|C\rangle_{\rm Gepner} = {\cal P} \left(\prod_{i=1}^r |C_i: NSNS\rangle
+ \prod_{i=1}^r |C_i: RR\rangle\right)
$$
${\cal P}$ imposes the $U(1)$ charge projection of Gepner. This is
the crosscap analogue of the Recknagel-Schomerus construction for boundary
states in the Gepner model\cite{RS}.

Now the crosscap state clearly carries RR charge.
It has all the terms to provide a supersymmetric
KB amplitude.
For the quintic, in fact, we find a full family of 625 distinct crosscap
states in agreement with the 625 anti-holomorphic involutions.
More detailed checks such as the KB projection, tadpole cancellation
etc. for specific examples will be discussed in
the paper to appear soon\cite{SGJM}. A recent paper by A. Misra also
discusses a type IIA orientifold of a Calabi-Yau threefold\cite{Aalok}.


\begin{thebibliography}{99}
\bibitem{Joyce} D.~Joyce, 
%`Compact Riemannian manifolds with holonomy $G_2$ I and II''
J. Diff. Geom. {\bf 43} (1996) 291 and
J. Diff. Geom. {\bf 43} (1996) 329.
%%CITATION = JDGEA,43,291;%%
%%CITATION = JDGEA,43,329;%%
%
\bibitem{HM}
J.~A.~Harvey and G.~W.~Moore,
%``Superpotentials and membrane instantons,''
arXiv:hep-th/9907026.
%%CITATION = HEP-TH 9907026;%%
\bibitem{PP}
H.~Partouche and B.~Pioline,
%``Rolling among G(2) vacua,''
JHEP {\bf 0103} (2001) 005
[arXiv:hep-th/0011130].
%%CITATION = HEP-TH 0011130;%%
%
\bibitem{BBS}
K.~Becker, M.~Becker and A.~Strominger,
%``Five-branes, membranes and nonperturbative string theory,''
Nucl. Phys.   {\bf B456} (1995) 130
[arXiv:hep-th/9507158].
%%CITATION = HEP-TH 9507158;%%
\bibitem{SGJM} S.~Govindarajan and J.~Majumder (to appear).
%%CITATION = NONE;%%
%
\bibitem{PT}
G.~Papadopoulos and P.~K.~Townsend,
%``Compactification of D = 11 supergravity on spaces of exceptional
%holonomy,''
Phys. Lett.  {\bf B357} (1995) 300
[arXiv:hep-th/9506150].
%%CITATION = HEP-TH 9506150;%%
%
\bibitem{VW}
C.~Vafa and E.~Witten,
%``Dual string pairs with N = 1 and N = 2 supersymmetry in four dimensions,''
Nucl.\ Phys.\ Proc.\ Suppl.\  {\bf 46} (1996) 225
[arXiv:hep-th/9507050].
%%CITATION = HEP-TH 9507050;%%
\bibitem{SYZ}
A.~Strominger, S.~T.~Yau and E.~Zaslow,
%``Mirror symmetry is T-duality,''
Nucl.\ Phys.\ B {\bf 479} (1996) 243
[arXiv:hep-th/9606040].
%%CITATION = HEP-TH 9606040;%%
%
\bibitem{Ashoke} A.~Sen,
%``A note on enhanced gauge symmetries in M- and string theory,''
JHEP {\bf 9709} (1997) 001 [arXiv:hep-th/9707123].
%%CITATION = HEP-TH 9707123;%%
\bibitem{Jaydeep} J.~Majumder,
%``Type IIA orientifold limit of M-theory on compact Joyce 8-manifold of
%Spin(7)-holonomy,''
JHEP {\bf 0201} (2002) 048
[arXiv:hep-th/0109076].
%%CITATION = HEP-TH 0109076;%%
\bibitem{GM}
R.~Gopakumar and S.~Mukhi,
%``Orbifold and orientifold compactifications of F-theory and M-theory to six
% and four dimensions,''
Nucl.\ Phys.\ B {\bf 479} (1996) 260
[arXiv:hep-th/9607057].
%%CITATION = HEP-TH 9607057;%%
\bibitem{KM}
S.~Kachru and J.~McGreevy,
%``M-theory on manifolds of G(2) holonomy and type IIA orientifolds,''
JHEP {\bf 0106} (2001) 027
[arXiv:hep-th/0103223].
%%CITATION = HEP-TH 0103223;%%
%
\bibitem{PSS}
G.~Pradisi, A.~Sagnotti, Ya.~S.~Stanev,
%``Planar Duality in $SU(2)$ WZW Models'',
Phys. Lett. {\bf B354} (1995) 279
[arXiv:hep-th/9503207].\\
G.~Pradisi, A.~Sagnotti, Ya.~S.~Stanev,
%``The Open Descendants of Non-Diagonal $SU(2)$ WZW Models'',
Phys. Lett. {\bf B356} (1995) 230
[arXiv:hep-th/9506014].
%%CITATIONS= HEP-TH 9503207;%%
%%CITATIONS= HEP-TH 9506014;%%
%
\bibitem{Hikida}
Y.~Hikida,
%``Orientifolds of SU(2)/U(1) WZW models,''
JHEP {\bf 0211} (2002) 035
[arXiv:hep-th/0201175].
%%CITATION = HEP-TH 0201175;%%
%
\bibitem{BH}
I.~Brunner and K.~Hori,
%``Orientifolds and mirror symmetry,''
arXiv:hep-th/0303135.
%%CITATION = HEP-TH 0303135;%%
%
\bibitem{RS}
A.~Recknagel and V.~Schomerus,
%``D-branes in Gepner models,''
Nucl.\ Phys.\ B {\bf 531} (1998) 185
[arXiv:hep-th/9712186].
%%CITATION = HEP-TH 9712186;%%
\bibitem{Aalok}
A.~Misra, 
% ``On (Orientifold of) type IIA on a compact Calabi-Yau''
arXiv:hep-th/0304209.
%%CITATION = HEP-TH 0304209;%%
\end{thebibliography}
\end{document}